**Dual Charge-Density-Waves in Two-dimensional $DyTe_3$ and Their Distinct Impacts on Magneto-Transport Properties**

Shuvankar Gupta[#,*], Yasemin Ozbek[#], Olajumoke Oluwatobiloba Emmanuel[#], Maya Bostock, Johnathan Kowalski, Pengpeng Zhang[*], and Xianglin Ke

Department of Physics and Astronomy, Michigan State University, East Lansing, Michigan 48824-2320, USA

[#]: These authors contributed equally to this work.

[*]: Corresponding author: guptas40@msu.edu, zhangpe@msu.edu

**Abstract**

Charge-density-waves (CDWs) are macroscopic quantum states defined by periodic modulations in electronic charge density coupled with lattice distortions. Despite significant research efforts, the evolution of electromagnetic transport properties in the presence of CDWs remains largely unexplored. Here, we report the effects of two orthogonal CDWs ($CDW_1$ and $CDW_2$) coexisting at low-temperature in $DyTe_3$ on its electronic transport properties. Importantly, we find that while no clear magnetotransport anomaly is resolved across $CDW_1$ transition under the present experimental conditions, the emergence of $CDW_2$ drives a drastic enhancement in magnetoresistance and a multiband-governed nonlinear Hall response. These results demonstrate that $CDW_1$ and $CDW_2$ in $DyTe_3$ are fundamentally different and impact its electronic transport in qualitatively distinct ways, establishing $DyTe_3$ as a model system for understanding the transport consequences of multiple CDWs in quasi-two-dimensional materials.

## Introduction:

A charge-density-wave (CDW) phase is a fascinating macroscopic quantum state characterized by the periodic modulation of electronic charge density, accompanied by a corresponding distortion of the atomic lattice [1,2]. In recent years, substantial progress has been made in understanding the physics governing cooperative CDW states [3–8]. These states often coexist with unconventional forms of superconductivity, underscoring their significance in condensed matter physics [9,10]. Early investigations into CDW phenomena primarily focused on quasi-one-dimensional (1D) systems where strongly bound atomic chains are weakly connected by van der Waals forces [11,12]. Later, attention shifted to two-dimensional (2D) transition metal dichalcogenides (TMDs), where the impact of dimensionality—from bulk to monolayer—on CDW stability and electronic properties became a major focus [13,14].

Quasi-two-dimensional (2D) Van der Waals layered crystals exhibiting CDW formation represent a broader class of materials with intriguing physical properties [15–19]. Among these, the lanthanide tritellurides $RTe_3$, where R denotes rare-earth elements La - Tm, have emerged as a versatile platform for exploring CDW behavior [20–32]. Interestingly, the CDW transition in $RTe_3$ materials sensitively depends on the rare-earth composition [33–35]. The variation of the internal chemical pressure due to different ionic radii of R in $RTe_3$ series significantly impacts the density of states near the Fermi-surface, the CDW ordering and its transition temperatures. On the one hand, the lighter rare-earth element based $RTe_3$ (R = La - Gd) compounds exhibit a single unidirectional, incommensurate CDW with a wave vector ($q_1 \sim 2/7\ c^*$ with $c^* = 2\pi/c$) oriented along the $c$-axis, denoted as $CDW_1$ [24,30,35]. In contrast, $RTe_3$ compounds with heavier elements (Tb-Tm) display an additional secondary CDW at lower temperatures, termed as $CDW_2$ which is characterized by a wave vector ($q_2 \sim 1/3\ a^*$ with $a^* = 2\pi/a$) along the $a$-axis that is orthogonal to the $CDW_1$ wave vector $q_1$ [27]. On the other hand, the transition temperature of $CDW_1$ decreases with decreasing lattice parameters, whereas $CDW_2$ transition temperature follows an opposite trend, decreasing as the lattice parameters increase and eventually disappearing midway through the series [27].

The $CDW_2$ phase in heavier rare earth series has been unambiguously studied through various experimental techniques [27,34–40], yet whether $CDW_1$ and $CDW_2$ originate from the same underlying mechanism remains controversial. Based on X-ray diffraction and electrical resistivity studies, it was initially proposed that the low-temperature $CDW_2$ modulation is simply the perpendicular analogue of $CDW_1$, implying a common Fermi-surface nesting-driven instability [27]. However, subsequent inelastic X-ray scattering measurements on $DyTe_3$ challenged this unified picture [39]. It was found that while $DyTe_3$ exhibits pronounced phonon softening at both $CDW_1$ and $CDW_2$ wave vectors below the first $CDW_1$ transition temperature ($T_{CDW1}$), intriguingly, no phonon soft-mode feature emerges below the $CDW_2$ transition ($T_{CDW2}$) [39]. The strong phonon softening below $T_{CDW1}$ is consistent with a robust electron–phonon–coupled instability, whereas the absence of comparable softening below $T_{CDW2}$ suggests that $CDW_2$ may be fundamentally different from $CDW_1$ [39]. Intriguingly, very recent time-resolved ARPES studies on $ErTe_3$ reveal distinct recovery dynamics of its two CDWs, which also suggests different mechanisms of these two CDW orders with the $CDW_2$ state formed via a nucleation-like first-order phase transition in contrast to a second-order phase transition for the $CDW_1$ state [41]. These new findings raise an important open question: do these two CDW states distinctly influence electronic transport?

Here, we combine scanning tunnelling microscopy (STM) / spectroscopy (STS) and magnetotransport measurements on $DyTe_3$ single crystals to examine the relationship between the two CDWs states and the electronic transport. While no clear magnetotransport anomaly is resolved across the $CDW_1$ transition under the present experimental conditions, we find that the onset of $CDW_2$ leads to a large enhancement of magnetoresistance and a multiband-driven nonlinear Hall response. These results suggest that the $CDW_1$ and $CDW_2$ states with orthogonal

CDW wave vectors in $DyTe_3$ are not simply analogous, underscoring the need to treat the two states as fundamentally distinct phenomena within the $RTe_3$ family with heavier R.

**Results and Discussion:**

Single crystals of $DyTe_3$ were grown using the self-flux method. Detailed growth procedures and experimental characterization methods are described in the Supplemental Material [42]. $DyTe_3$ crystallizes in a space group #63 (*Cmcm*) with lattice parameters $a$ = 4.302 Å, $b$ = 25.43 Å, $c$ = 4.304 Å [30,43,44]. As illustrated in Figure 1(a), $DyTe_3$ has a layered structure, comprising two planar Te nets interleaved with buckled DyTe-slabs in the *ac*-plane. The layered structure with weak interlayer interactions leads to pronounced anisotropic physical properties [44]. The temperature-dependent magnetic susceptibility, measured under zero-field-cooled conditions with the magnetic field applied along the crystallographic $a$, $b$, and $c$-axes, reveal a sharp anomaly at $T_N \sim 4.5$ K, as shown in Figure 1(b), which corresponds to paramagnetic-to-antiferromagnetic phase transition that is consistent with the literature [26,44]. At low-temperatures, the magnetic susceptibility ($\chi$) shows its highest enhancement when the magnetic field ($H$) is oriented along the $c$-axis, aligning with findings from recent studies [44]. The isothermal magnetization as a function of the applied magnetic field ($M$ vs. $H$) was measured at various temperatures, as shown in Figure 1(c) for $T = 2$ K and Figure S1 [42] for other temperatures. The field-induced metamagnetic transitions below $T_N$ align with the recent observation of the helimagnetic, cone-type ordering of $DyTe_3$ using polarized elastic neutron scattering [44]. The anisotropic response of magnetization data along with isothermal magnetization along the $a$, $b$, and $c$-axes confirms the strong directional dependence of the magnetic interactions, which can be attributed to the underlying crystalline and magnetic structure of $DyTe_3$.

First, we use STM and STS to probe the $CDW_1$ and $CDW_2$ states. Based on the temperature dependence of resistivity measurements, it was reported that $DyTe_3$ undergoes two CDW phase transitions, with $CDW_1$ occurring at $T_{CDW1}$ ~ 308 K followed by the emergence of $CDW_2$ at $T_{CDW2}$ ~ 50 K which coexists with $CDW_1$ [27]. We have performed STM and STS measurements at 77 K ($T_{CDW1} > T > T_{CDW2}$, Figures 2(a-c)) and 4.5 K ($T < T_{CDW2}$, Figures 2(d-f)). Figures 2(a,d) and 2(b,e) show the STM topography image and the corresponding fast Fourier transform (FFT) of the image, respectively. The blue lines in Figure 2(a) serve as a guide to the eyes. Figure 2(b) marks the block lattice FFT peaks in blue squares and the Te lattice peaks in red squares. A primary incommensurate CDW along the *c*-direction can also be seen. The wavevector of this CDW is analysed based on the FFT, where the purple-colored circles in Figures 2(b,e) illustrate the Bragg peaks associated with the $CDW_1$. Line cut along the *c** direction shows several pronounced peaks located at ~2/7, ~3/7, ~4/7, and ~5/7 of *c**, as seen in Figure S2 [42]. The primary wavevector of the incommensurate $CDW_1$ is ~2/7 *c**, with other Bragg peaks appearing at $c^* - 2q_{cdw}$ (i.e., ~3/7 *c**), $2q_{cdw}$ (i.e., ~4/7 *c**), and $c^* - q_{cdw}$ (i.e., ~5/7 *c**) which are related to the second-order harmonic lattice distortions and the corresponding equivalent peaks. In contrast, with the sample cooled down to 4.5 K, the STM image clearly exhibits a checkerboard pattern arising from two coexisting CDW phases, as shown in Fig. 2(d). Detailed analysis of the FFT, presented in Fig. 2(e), demonstrates that $DyTe_3$ exhibits a $CDW_2$ which has a wavevector of ~ 1/3 *a** along *a**. The ~1/3 a* and ~2/3 a* wavevectors are circled in yellow in Fig.2(e), together with the incommensurate $CDW_1$ that shows the same wavevectors of ~2/7 *c**, ~3/7 c*, etc as that observed at 77 K. That is, $CDW_1$ and $CDW_2$ coexist at low-temperature with their wavevectors orthogonal to each other. These observations are consistent with a very recent STM study [40].

Figure 2(c) and 2(f) present the STS of $DyTe_3$ measured at 77 K and 4.5 K, respectively. The obtained dI/dV curves provide useful information on the electronic structure with and without $CDW_2$. The signal is directly proportional to the density of states, which allows us to identify the spectra features related to the formation of each CDW. By comparing the angle-resolved photoemission spectroscopy (ARPES) and STM/STS studies on other $RTe_3$ materials [28,31,34], we can assign the outer peaks at -0.214 V and 0.219 V of the dI/dV curve shown in Figure 2(f) to the $CDW_1$ phase and the inner two peaks at -0.060 V and 0.070 V to the $CDW_2$ phase. The gap centre is not located exactly at the Fermi energy, which is consistent with the ARPES observation in other $RTe_3$ compounds [34]. An outline of the CDW peak analysis can be seen in Figure S3 [42]. At 77 K, the signal of the inner two peaks is not as pronounced as that measured at ~ 4.5 K (Fig. 2(f)), which agrees with the existence of the weak $CDW_2$ phase. This suggests that in $DyTe_3$ the $CDW_1$ phase dominates at 77 K, but there may exist a possible $CDW_2$-related precursor, implying the possible precursor of the onset of long-range correlated $CDW_2$ phase at $T_{CDW2}$. Recent inelastic X-ray scattering measurements suggest that the second $CDW_2$ transition in $DyTe_3$ occurs near 68 K [39], which may be associated with the development of short-ranged correlated $CDW_2$ phase, in contrast to the 50 K transition temperature inferred from resistivity data [27]. The morphology and spectral differences observed between 4.5 K and 77 K in the STM/STS measurements, as discussed above, provide compelling evidence for the emergence of a stable $CDW_2$ phase in $DyTe_3$ at low-temperature.

Next, we present the effects of $CDW_2$ on the electronic- and magneto-transport properties of $DyTe_3$, which were measured on a different piece of single crystal that was grown in the same batch as the crystal used for STM/STS studies. The temperature-dependent resistivity ($\rho$) of $DyTe_3$, as presented in Figure 3(a), shows characteristic $\rho(T)$ features of the $RTe_3$ family [27]. At zero magnetic field, a high residual-resistivity-ratio (RRR), which is defined as $\rho$ (300 K) / $\rho$ (2 K), reaches a value of 47, indicating high single crystal quality. The RRR of 47 reported here is comparable to other reported values of 26 [45] and 47 [46] ,

reflecting the good sample quality, which is supported by the STM and transport measurements that capture the intrinsic electronic properties of $DyTe_3$. Around room temperature, $DyTe_3$ exhibits semiconducting-like behavior, which is ascribed to the electronic gap opening below the $CDW_1$ phase transition temperature (at ~ 308 K). Below ~ 280 K $DyTe_3$ exhibits metallic-like behavior with the resistivity decreasing as the temperature decreases, a phenomenon attributed to partial gap formation [30]. Interestingly, unlike the hump in $\rho(T)$ observed near $T_{CDW1}$, $\rho(T)$ does not show an obvious anomaly near $T_{CDW2}$ ~ 50 K, which is in an agreement with the previous report [27]. Nevertheless, a slope change (decrease) in the plot of $d\rho/dT$ versus temperature for 0 T is observed near $T_{CDW2}$ (Figure 3(a)), implying a potential alteration of electronic structure near the Fermi level due to the occurrence of the second CDW ordering, which is consistent with the STM measurements at zero magnetic field. These results align well with the observations reported in previous studies [27,45].

Figures 3(b) and 3(c) display the magnetic field-dependence of magnetoresistance (MR) for $H \parallel b$-axis and $H \parallel ac$-plane, respectively, with current applied along the $a$-axis. In both orientations, MR is large, positive, and non-saturating, reaching ~600% ($H \parallel b$-axis) and ~ 200% ($H \parallel ac$-plane) at 2 K. Notably, this behavior is not primarily associated with the onset of long-range antiferromagnetic order, as it persists both below and above the Néel temperature ($T_N$ ~ 4.5 K). Scattering from paramagnetic Dy 4*f* moments or short-range magnetic correlations above $T_N$ cannot be completely excluded; however, the susceptibility shows no magnetic transition near 50 K [26,44], whereas the pronounced changes in the MR and Hall response emerge in the same temperature range as $T_{CDW2}$. The non-saturating high-field magnetoresistance observed up to 5 T is not associated with any magnetic phase transition, as confirmed by the magnetization measurements (Figure 1 (c)), which show no indication of metamagnetic phase transition behaviour up to this field. At low fields, MR is nearly linear, but deviations emerge at higher fields, with stronger anisotropy for $H \parallel ac$-plane. The observed direction-dependent magnetoresistance behavior arises from the reconstructed Fermi-surface associated with the CDW order. A striking feature is the pronounced enhancement of MR below ~50 K (Fig. 3(d)), coinciding precisely with the onset of the $CDW_2$. This correlation directly connects the formation of $CDW_2$ with the unusual magnetotransport properties. While MR values are large across the $RTe_3$ family [47–50], $DyTe_3$ shows strong evidence that $CDW_2$ plays a central role. The bifurcation of $\rho(T)$ measured with and without magnetic field (Fig. 3a) further demonstrates the influence of $CDW_2$ on electronic transport. To further support this argument, we also measured $ErTe_3$, where $CDW_2$ occurs at $T_{CDW2} \sim 160$ K. As shown in Fig. S4 [42], the resistivity $\rho(T)$ curves measured at 0 T and 5 T deviate below $T_{CDW2}$, a feature similar to the $\rho(T)$ data of $DyTe_3$. These results indicate that the large MR enhancement at low-temperature is intrinsically tied to the emergence of $CDW_2$ and is therefore a universal feature across the $RTe_3$ series. In contrast, we find that the onset of $CDW_1$ has little effect on the magnetotransport properties. We can see that $\rho(T)$ shows no discernible difference between 0 and 3 T data (figure 3 (a)). Additionally, Figure S5 [42] shows the temperature-dependent resistivity of another $DyTe_3$ sample measured at both 0 T as well as the MR and Hall measurements at various temperatures. We can see the MR and Hall signal are very small and nearly unchanged across $T_{CDW1} \sim 308$ K. The high-temperature measurements in Figure S5 [42] were performed on a crystal from a different growth batch with RRR $\approx$ 20, compared with RRR $\approx$ 47 for the main low-temperature sample. In these measurements, no clear anomaly in the MR or Hall response was resolved across $T_{CDW1}$ under the present experimental conditions. This contrasts with the trend of $\rho(T)$ and MR features below $T_{CDW2}$, which are consistent with the data shown in Figure 3(a, b, d).

The large, non-saturating MR below $T_{CDW2}$ in $DyTe_3$ prompts an examination of its microscopic origin. One possibility is electron–hole compensation, where nearly equal carrier concentrations enhance MR through Hall effects, as seen in $WTe_2$ [51,52]. Although spectroscopy has revealed comparable electron and hole pockets in $RTe_3$ [53], the modest anisotropy in MR for $DyTe_3$ (($H$ // *b-axis*) / ($H$ // *ac*-plane) ≈ 3) is inconsistent with compensation-driven systems, where anisotropy typically exceeds $10^2$. Another scenario is the extreme quantum limit, in which the cyclotron energy exceeds the Fermi energy and electrons collapse into the lowest Landau level, producing linear MR [54,55]. $DyTe_3$, with its small Fermi pockets and low effective masses due to the square-net band structure [56–58], could be a candidate. Another prominent explanation is the Fermi-surface reconstruction induced by $CDW_2$. The additional periodic modulation associated with $CDW_2$ creates anisotropic "hot spots" on the Fermi-surface, enhancing quasiparticle scattering as previously proposed for $TbTe_3$ and $HoTe_3$ [50]. In particular, magnetic breakdown near the boundaries between gapped and ungapped Fermi-surface regions enhances quasiparticle scattering, which strengthens the field-dependence of charge transport [50]. In $DyTe_3$, where $CDW_2$ forms intrinsically at zero field, the large MR enhancement below $T_{CDW2}$ and the bifurcation of resistivity under field provide direct evidence of this connection.

Within this picture, electron–hole compensation may be viewed as a secondary consequence of the Fermi-surface reconstruction driven by $CDW_2$, rather than the primary origin of the large magnetoresistance. While partial compensation can further enhance MR, the dominant contribution arises from the strong reconstruction of the Fermi-surface, which generates small, high-mobility pockets and pronounced momentum-dependent scattering ("hot spots") [50]. These features strongly enhance the sensitivity of electronic transport to magnetic field and naturally account for the large, non-saturating MR observed below $T_{CDW2}$. One may ask why lighter rare-earth members of the $RTe_3$ family also exhibit large MR at low-temperatures despite the absence of an intrinsic $CDW_2$ transition in the absence of applied magnetic field. In this regard, it is worth noting that $LaTe_3$ has been proposed to host a field-induced $CDW_2$ state at low-temperatures [47], while $CeTe_3$—long considered a single-CDW

system—was recently shown by STM/STS to develop two additional CDW states, including $CDW_2$ [59], under an in-plane magnetic field. In contrast, $DyTe_3$ naturally develops $CDW_2$ at zero field, as confirmed by our STM/STS measurements (Fig. 2(d–f)), making it a unique platform to temperature-dependent $CDW_2$ formation with the enhanced magnetotransport. These results establish a direct experimental correlation between the intrinsic formation of $CDW_2$ and enhanced magneto-transport in $DyTe_3$ and suggest that the second $CDW_2$ plays an important role in related $RTe_3$ compounds.

The formation of the $CDW_2$ state also has a profound impact on the Hall response of $DyTe_3$. As shown in Fig. 4(a), the Hall resistivity ($\rho_H$) is linear and positive for temperatures between $T_{CDW1} > T > T_{CDW2}$. Strikingly, below $T_{CDW2}$, $\rho_H$ becomes nonlinear and changes sign, turning negative with increasing field. This crossover, highlighted in Fig. 4(b) where $\rho_H$ at 5 T switches from positive to negative near 50 K, signals a drastic reorganization of the electronic structure. Such behavior reflects multi-band transport driven by Fermi-surface reconstruction, a hallmark of CDW formation [60]. Complementary measurements on $ErTe_3$ (Fig. S6 [42]) reveal an analogous sign reversal of $\rho_H$ near its $CDW_2$ transition at 160 K, while similar features in $LaTe_3$, $NdTe_3$, and $TbTe_3$ and very recent report on $DyTe_3$ [47–50,61] were reported earlier but without attributing them to $CDW_2$. Our results identify $CDW_2$ as a key ingredient governing the Hall response in $DyTe_3$ and provide an experimental basis for re-examining similar Hall anomalies reported in other $RTe_3$ compounds.

To parameterize the evolution of the Hall response, we simultaneously fitted ρ(B) and $\rho_H$(B) using the minimal two-band model [62,63] of equations (1) and (2). The fitted quantities are denoted as effective carrier densities ($n_h$ and $n_e$) and effective mobilities ($\mu_h$ and $\mu_e$) because each parameter represents the aggregate contribution of electron-like or hole-like carriers and cannot be assigned uniquely to a single Fermi pocket. The model captures the overall field-dependence, whereas systematic deviations become more evident below $T_{CDW2}$, indicating that more than two carrier channels may contribute to the reconstructed state. We therefore use these effective parameters only to track the temperature evolution of the multiband transport. The fitting procedure, residuals, and parameter uncertainties are provided in Figure S7 [42] and Table S3 [42].

$$\rho = \frac{1}{e} \cdot \frac{(n_e \mu_e + n_h \mu_h) + (n_e \mu_h + n_h \mu_e) \mu_e \mu_h B^2}{(n_e \mu_e + n_h \mu_h)^2 + (n_e - n_h)^2 \mu_e^2 \mu_h^2 B^2}, \qquad (1)$$

$$\rho_H = \frac{B}{e} \cdot \frac{(n_h \mu_h^2 - n_e \mu_e^2) + (n_e - n_h) \mu_e^2 \mu_h^2 B^2}{(n_e \mu_e + n_h \mu_h)^2 + (n_e - n_h)^2 \mu_e^2 \mu_h^2 B^2}, \qquad (2)$$

The extracted effective carrier mobilities ($\mu_h$ and $\mu_e$) and carrier concentrations ($n_h$ and $n_e$) for holes and electrons are presented in Figures 4(c) and 4(d), respectively. As can be shown in Figure S7 [42], while the fits reproduce the data well for $T > 50$ K, deviations at lower temperatures suggest the involvement of additional bands, consistent with ARPES reports of complex Fermi-surface reconstruction in $RTe_3$ [56,57]. Similar deviations have also been reported in other $RTe_3$ compounds (R = La, Ce, Tb), suggesting a broader trend within this material family [48]. The reconstructed Fermi-surface introduced by $CDW_2$ not only generates anisotropic scattering but also boosts carrier mobility, a trend observed throughout the $RTe_3$ family [47–50,58], but not explicitly linked to CDW order. Importantly, below $T_{CDW2}$, both carrier mobilities and densities increase drastically, underscoring that $CDW_2$ not only reconstructs the Fermi-surface but also enhances the effectiveness of charge transport.

Intriguingly, in contrast to the intuitive expectation that the carrier concentration decreases below the onset of CDW order due to Fermi-surface gapping, $DyTe_3$ exhibits an increase in both electron and hole effective concentrations below $T_{CDW2.}$ Within the minimal two-band representation, this behavior reflects a pronounced Fermi-surface reconstruction induced by the second CDW order, which generates multiple small electron and hole pockets rather than simply removing states at the Fermi level. Such a unique character may stem from the existence of two orthogonal CDWs in $DyTe_3$ instead of a single CDW as found in conventional CDW materials such as $TiSe_2$ [60]. In conventional CDW systems, the formation of an energy gap on nested portions of the Fermi-surface tends to reduce the density of states at the Fermi level and the number of carriers available for conduction. Recently, a quantum oscillation study on heavier $RTe_3$ compounds exhibiting two CDW states reveals that the emergence of the second CDW state at low-temperatures leads to the formation of multiple small, light electron and hole pockets [64]. ARPES studies on other $RTe_3$ compounds [30,31,57] along with theoretical calculations on $LaTe_3$ [47] have shown that the field-induced $CDW_2$—significantly reconstructs the Fermi-surface, modifying the electronic structure and enhancing the formation of these carrier pockets. Notably, DFT+DMFT simulations for $LaTe_3$ [47] demonstrated that this Fermi-surface reconstruction arises from coupling to in-plane and out-plane external magnetic field and results in enhanced carrier mobility and magnetoresistance. While $LaTe_3$ [47] requires an external magnetic field to stabilize the second CDW order, $DyTe_3$ intrinsically develops the $CDW_2$ phase at zero field due to its smaller lattice constants and larger internal chemical pressure. Our observation of increased carrier concentrations and enhanced magnetoresistance below $T_{CDW2}$ is consistent with this theoretical framework of $LaTe_3$ [47], suggesting that a similar Fermi-surface reconstruction mechanism operates in $DyTe_3$. This raises an intriguing question: how does the formation of additional pockets due to the $CDW_2$ order give rise to an increase in carrier concentration. While further theoretical investigations are highly desirable to understand the underlying mechanisms, our experimental observation offers solid evidence of the Fermi-surface reconstruction associated with the emergence of the $CDW_2$ that dramatically impacts the electronic transport properties in $DyTe_3$ and other $RTe_3$ compounds in general.

The observation of the drastically different effects of $CDW_1$ and $CDW_2$ orders in $DyTe_3$ on the magnetotransport properties, including the MR effect and Hall transport, suggests distinct mechanisms of these two charge orders. This finding is in line with the very recent report of different recovery dynamics of the two CDWs in $ErTe_3$ [41]. Intuitively it can be understood in terms of how the two CDW states reconstruct the Fermi-surface. Although $CDW_1$ gaps a substantial fraction of the electronic states, the associated Fermi-surface reconstruction does not appear to generate small electron and hole pockets and therefore produces only a weak response in magnetoresistance and Hall transport. In contrast, the low-temperature $CDW_2$ develops on an already partially gapped Fermi-surface and further reconstructs the remaining electronic states into small electron and hole pockets. These pockets are associated with high-mobility carriers and exhibit pronounced momentum-dependent scattering (“hot spots”) induced by the additional $CDW_2$ modulation. Such an electronic structure is inherently more susceptible to an applied magnetic field, providing a natural explanation for the emergence of large magnetoresistance and nonlinear Hall response only below $T_{CDW2}$. This is consistent with the ARPES [57] and quantum oscillation [58,64] studies and further supports the view that $CDW_1$ and $CDW_2$ have qualitatively different impacts on the low-energy electronic structure in $DyTe_3$. Future theoretical studies are required to clarify the mechanisms and nature of the two CDWs in heavier $RTe_3$ compounds. Beyond fundamental interest, the intrinsic coupling between multiple CDWs and magnetotransport in $DyTe_3$ offers a controllable platform for emergent functionalities in two-dimensional materials. The pronounced field-tuneable resistivity and Hall response below $T_{CDW2}$ indicate that charge-order reconstruction can serve as an efficient electronic control knob, with possible applications in anisotropic sensors, low-power memory elements, and switchable quantum devices.

We emphasize that the present work does not attempt to determine the complete momentum-resolved Fermi-surface of $DyTe_3$, which would require ARPES measurements or electronic-structure calculations including the low-temperature $CDW_2$ modulation. Instead, our conclusion is based on a set of mutually consistent experimental observations that constrain the possible origin of the magneto-transport anomalies. First, STM/STS directly confirms the emergence of the low-temperature $CDW_2$ state in $DyTe_3$, together with additional spectral features associated with the second CDW gap. Second, the enhancement of magnetoresistance and the onset of nonlinear, sign-changing Hall response occur only below $T_{CDW2}$, while no comparable change under present experimental condition is observed across the $CDW_1$-dominated temperature regime. Third, these anomalies occur far above the antiferromagnetic ordering temperature of $DyTe_3$, excluding magnetic ordering as their primary origin. Fourth, the two-band analysis shows a sharp increase in both mobility and effective carrier density below $T_{CDW2}$, indicating that the second CDW does not simply remove carriers through gap opening, but instead reorganizes the low-energy electronic states into multiple conducting pockets. Taken together, these observations support a $CDW_2$-induced electronic reconstruction scenario. While the detailed Dy-derived electronic structure may influence the precise reconstructed band topology, the dominant low-energy conducting states in $RTe_3$ are associated with the Te square-net bands [30]. Therefore, comparisons with previous ARPES [57], quantum-oscillation [58,64], and theoretical studies [47] on related $RTe_3$ compounds provide a physically reasonable framework, although future $DyTe_3$-specific calculations will be important for quantitatively resolving the reconstructed Fermi-surface.

**Conclusion:**

In summary, this study presents a comprehensive investigation into the CDW phenomena in two-dimensional $DyTe_3$, a rare-earth tritelluride that bridges the lighter $RTe_3$ members exhibiting a single CDW with the heavier $RTe_3$ members exhibiting two orthogonal CDWs. We show presence of two CDWs in $DyTe_3$ through detailed STM and STS measurements. Furthermore, we demonstrate that $DyTe_3$ exhibits a significant enhancement in magnetoresistance and a multi-band-driven Hall effect below the $CDW_2$ phase transition, whereas $CDW_1$ produces no measurable change in either magnetoresistance or Hall response within the limitation of the present measurements. This stark contrast indicates that $CDW_1$ and $CDW_2$ are likely governed by fundamentally different microscopic mechanisms and gap different regions of the Fermi-surface, leading to fundamentally different consequences for charge transport. These results highlight $DyTe_3$ as a model system to study the coexistence of multiple CDWs and their impacts on magneto-transport properties in $RTe_3$ compounds, which offers new insights on the investigation of multiple-CDWs related magneto-transport phenomena in other quasi-two-dimensional van der Waals materials.

**Acknowledgements**

S. G., O.E. and X.K. acknowledge financial support from National Science Foundation (DMR-2219046) and the Jenison Fund from Michigan State University. The electronic transport measurements were supported by the U.S. Department of Energy, Office of Science, Office of Basic Energy Sciences, Materials Sciences and Engineering Division under Grant No. DE-SC0019259. STM experiments were supported by the U.S. Department of Energy (DOE), Office of Basic Energy Sciences, Division of Materials Sciences and Engineering under Award Number DE-SC0019120. P. P. Zhang acknowledges the financial support from National Science Foundation (DMR-2112691).

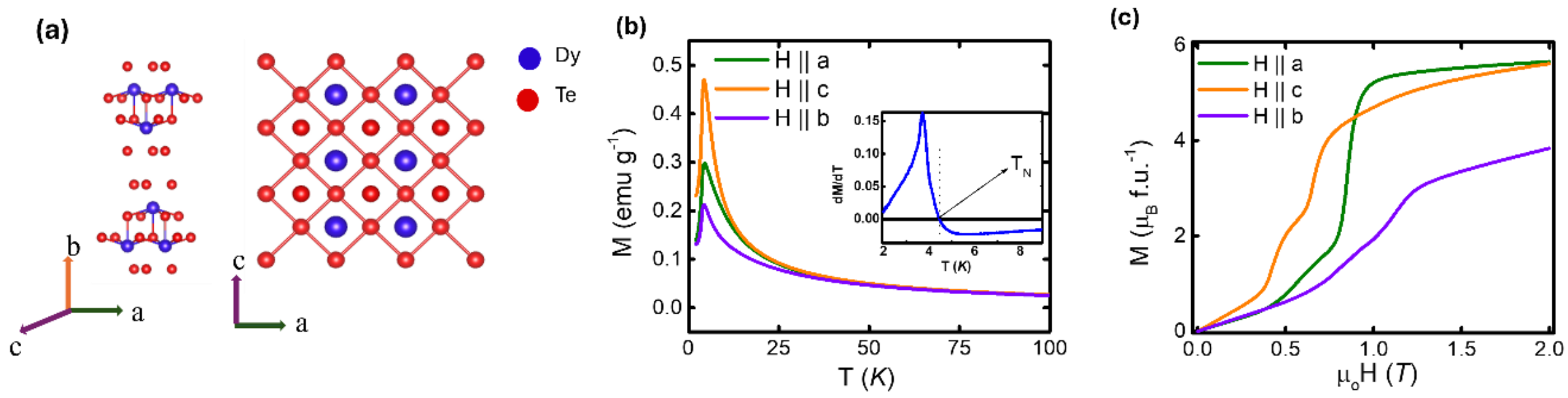

Figure 1: (a) Crystal structure of $DyTe_3$, showing the overall structure (left) and an in-plane (ac-plane) view highlighting the square Te-planes and rare earth block layer. (b) Temperature dependence of magnetization (M) along different crystallographic directions. (c) Isothermal magnetization measured at 2 K for various crystallographic directions.

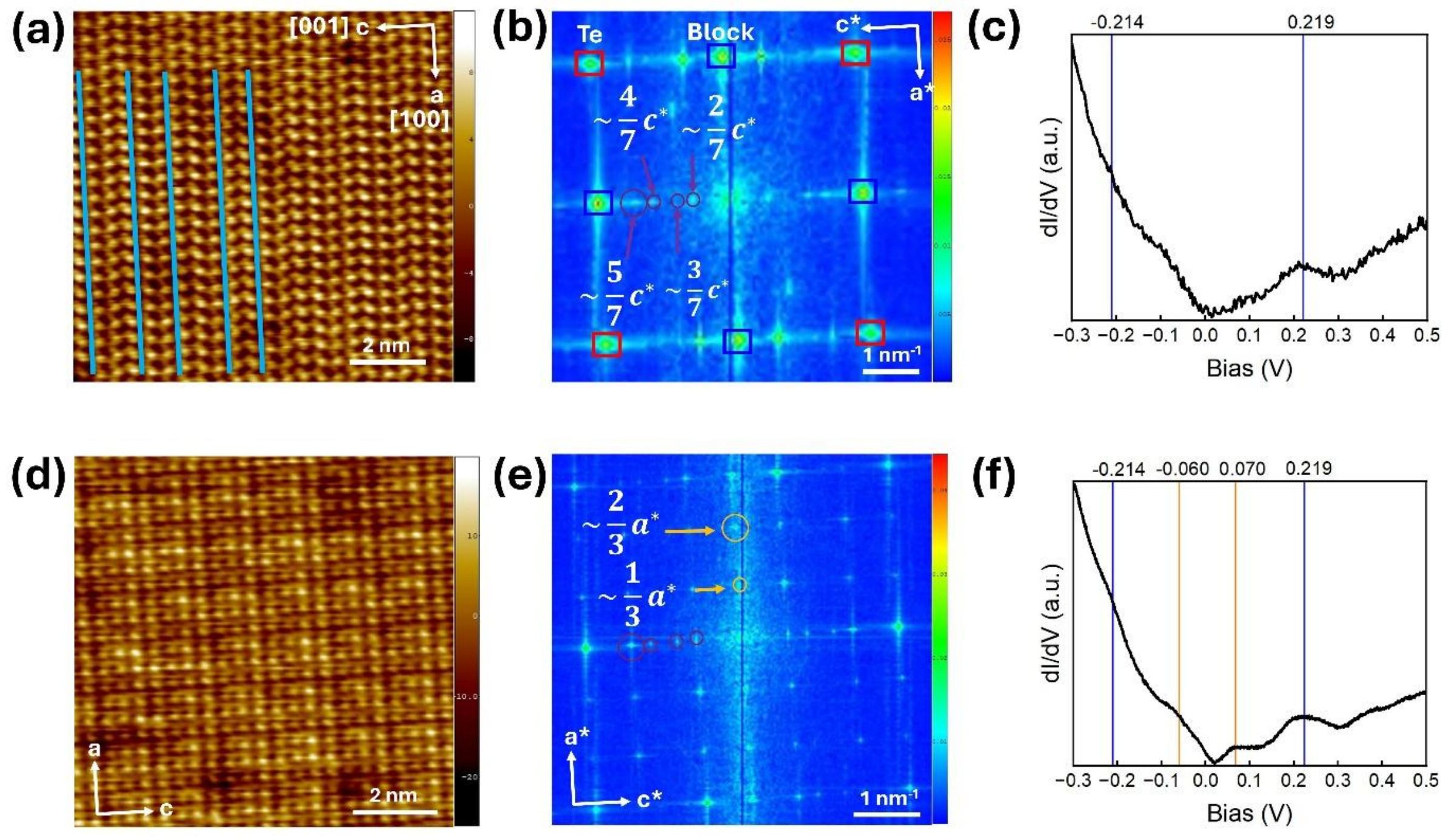

Figure 2: (a-c) STM data at 77K. (a) Morphology scan displaying the incommensurate first CDW phase (Vs = -0.05V, It= 300pA) along the light blue lines. (b) Fast Fourier Transform (FFT) of the image in (a) exhibiting the four block lattice peaks (blue squares), four Te lattice peaks (red squares), and in purple circles all four CDW 1 peaks at ~2/7 $c^*$, ~3/7 $c^*$, ~4/7 $c^*$, and ~5/7 $c^*$. (c) Average STS spectrum of 25 individual curves taken on different spots on the bulk sample away from defects (Vs = -0.5V, It = 50pA). Vertical blue lines represent the gap for the first CDW. (d-f) STM data at 4.5 K. (d) Morphology scan displaying both the incommensurate first CDW phase (running horizontally across the image) and incommensurate second CDW phase (running vertically across the image) (Vs = 0.5V, It = 50pA). (e) FFT of the image in (d) clearly showing all the peaks in (b) plus the ~1/3 $a^*$ and ~2/3 $a^*$ spots for CDW2, highlighted by the yellow circles. (f) Average STS spectrum of 6 individual curves taken on different spots on the bulk sample away from defects (Vs = 0.35V, It = 100pA). The blue lines correspond to the same gap as in (c). The orange lines represent the CDW2 gap. (Note: The STM images shown in Figure 2(a, d) were obtained on the same sample but at different fields of view (FOV). The difference in FOV arises from the necessary retraction and re-approach of the STM tip during temperature changes. These images represent prototype topographies recorded at 77 K and 4.5 K, respectively.)

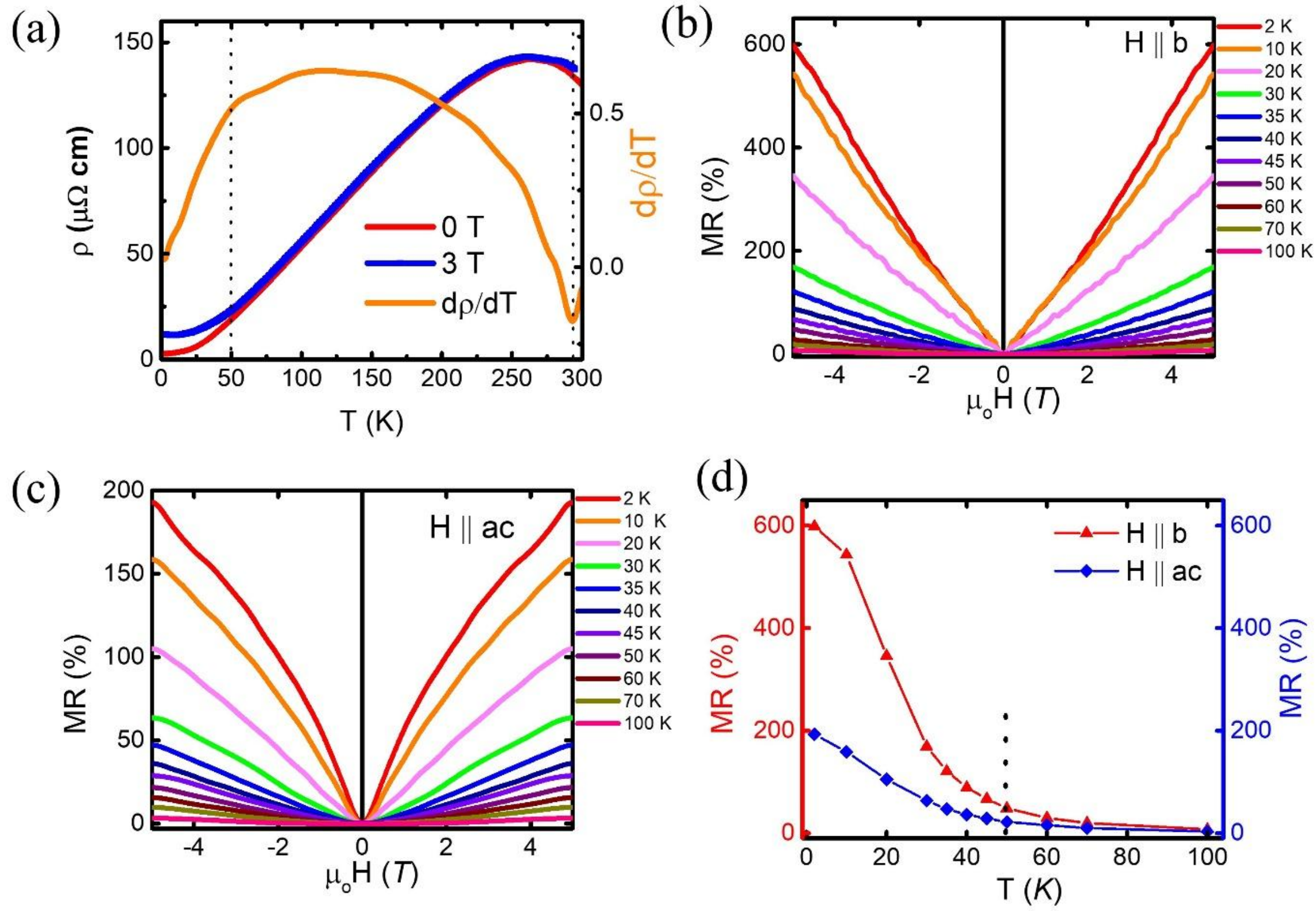

Figure 3: (a) Temperature dependence of in-plane (ac-plane) resistivity measured at 0 T and 3 T (ρ, left axis) and its first derivative with temperature (dρ/dT) for 0T (right axis). The vertical line near 50 K highlights anomalies in dρ/dT associated with the onset of the second charge-density-wave ($CDW_2$). (b) Magnetic field-dependence of in-plane magnetoresistance (MR) measured at various temperatures, with the magnetic field applied along the *b*-axis, perpendicular to the *ac*-plane. (c) Magnetic field-dependence of in-plane MR measured at various temperatures, with the magnetic field applied along the ac-plane direction. (d) Temperature dependence of MR for both field directions (H || *b*-axis and H || *ac*-plane). The

vertical line at 50 K corresponds to the emergence of the second CDW ($CDW_2$), marked by an increase in MR for both directions below this temperature.

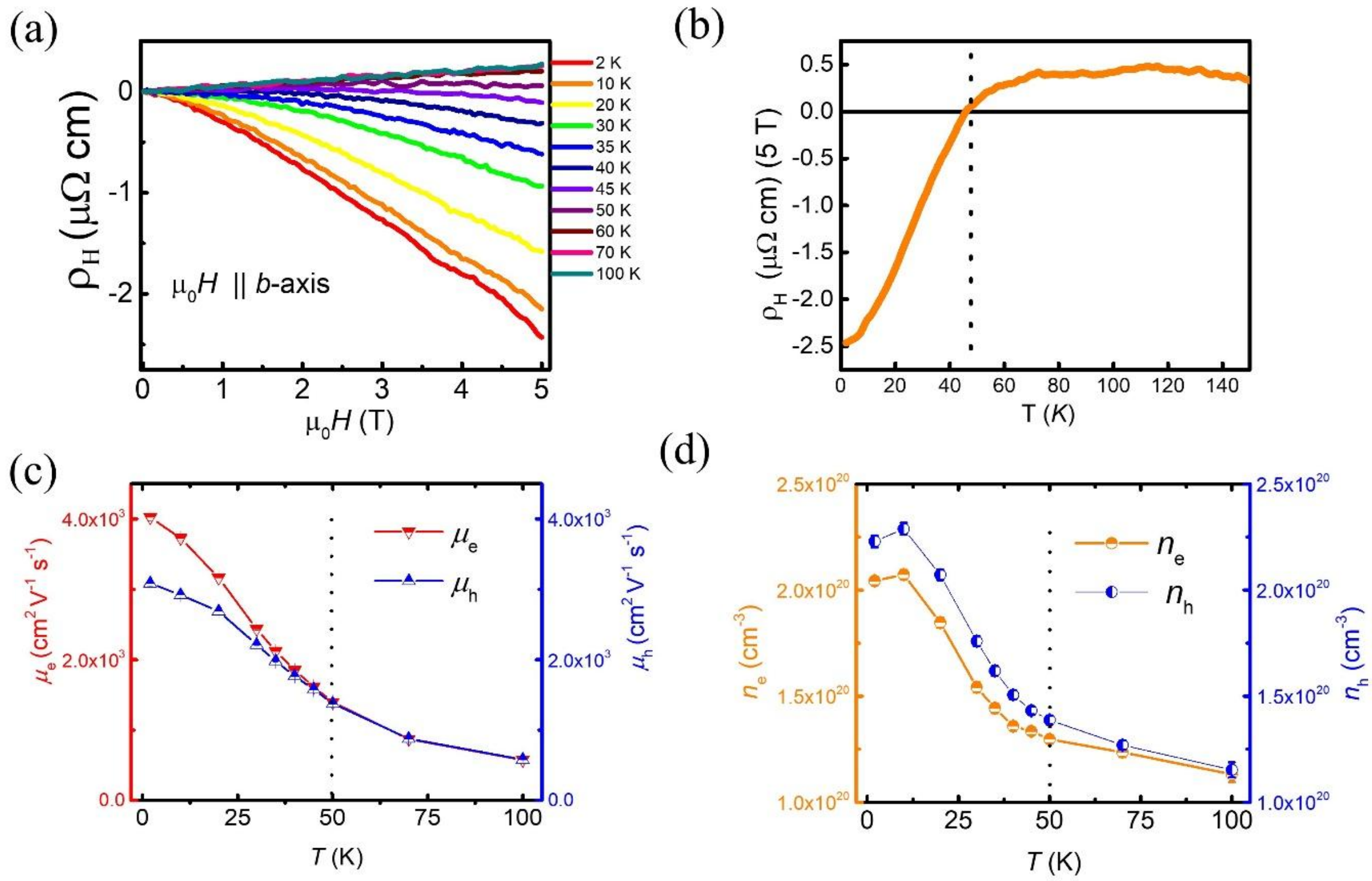

Figure 4: (a) Magnetic field-dependence of in-plane (*ac*-plane) Hall resistivity ($\rho_H$) measured at various temperatures, with the magnetic field applied along the *b*-axis, perpendicular to the ac-plane. (b) Temperature dependence of $\rho_H$ measured at 5 T. The vertical line near 50 K marks the crossover from a positive to a negative region on colling, corresponding to the emergence of the $CDW_2$. (c) Temperature dependence of effective electron and hole mobilities, derived from two-band model fittings. (d) Temperature dependence of electron and hole carrier effective concentrations. Error bars (smaller than the symbol size) are included for all data points. The fitting residuals are < 3%, confirming excellent agreement between the experimental data and the two-band model. The simultaneous fits and field-dependent residuals are shown in Fig. S7 [42].